\documentclass[
    ,final            
  ]
  {aipproc}

\layoutstyle{6x9}

\begin{document}

\title{Gaseous Debris Disks around White Dwarfs}


\author{Boris T. G\"ansicke}{
  address={University of Warwick, Department of Physics, Gibbet Hill
  Road, Coventry CV4 7AL, UK}
}

\begin{abstract}
This is a short and rather narrative summary of our ongoing efforts in
identifying white dwarfs with gaseous debris disks, and developing an
understanding of the structure and origin of these disks.
\end{abstract}

\maketitle


\section{A serendipitous discovery}
Early in 2006, while inspecting many thousand spectra of white dwarfs
from the Sloan Digital Sky Survey (SDSS, \citealt{yorketal00-1,
adelman-mccarthyetal08-1, abazajianetal09-1}) as part of a large
project on white dwarfs with low-mass
companions \citep{rebassa-mansergasetal07-1}, we noticed emission
lines of the Ca\,{\sc ii} $I$-band triplet in the spectrum of
SDSS\,J122859.93+104032.9 (Fig.\,1). On closer inspection, these lines
exhibited a double-peaked shape that is the hallmark of gas rotating
within an accretion disk \citep{horne+marsh86-1}. In fact, the full
width of the lines, and the separation of their double-peaks is very
similar to those observed in many cataclysmic variables (CVs),
i.e.\ white dwarfs accreting from a nearby companion star (see
e.g.\ Fig.\,1 in \citealt{southworthetal06-1}). At optical wavelengths,
no trace of a potential companion star to SDSS\,J1228+1040 is seen,
leaving only the possibility of a sub-stellar companion, such as found
in a handful of CVs \citep{littlefairetal06-1,
littlefairetal08-1}. The most striking difference to any known CV was,
however, the absence of Balmer or He emission lines. Time-resolved
follow-up observations obtained on the William Herschel Telescope
(WHT) failed to reveal any significant radial velocity variation of
the white dwarf on time scales of hours to days, corroborating the
hypothesis that SDSS\,J1228+1040 is a \textit{single} white dwarf with
a circumstellar gaseous disk depleted in volatile elements. The
detection of photospheric Mg absorption also demonstrated that the
white dwarf is accreting from the circumstellar material.

Pondering about the possible origin of the circumstellar disk around
SDSS\,J1228+ 1040, memories of \textit{two} ApJ Letters published
back-to-back, announcing the detection of dusty debris disks around
the white dwarf GD\,362 flooded back (\citealt{becklinetal05-1,
kilicetal05-1}, see Farihi, \nocite{farihi2010} this volume, 
for a review on debris disks
around white dwarfs) and the link was established: we had found a
white dwarf with a gaseous debris disk, and the dynamics of the
double-peaked line profiles unambiguously showed that the material was
residing within the tidal disruption radius of the white
dwarf \citep{gaensickeetal06-3}. Subsequent observations
with \textit{Spitzer} revealed that SDSS\,J1228+1040 also exhibits an
infrared flux excess, demonstrating the additional presence of a dusty
debris disk component \citep{brinkworthetal09-1}.

\section{And then there were three}
The natural consequence of such an exciting discovery was to search
for more objects of this kind. Scrutinising Data Release (DR)\,5
revealed a second DA white dwarf with Ca\,{\sc ii} emission lines,
SDSS\,J104341.53+085558.2 \citep{gaensickeetal07-1}. As in
SDSS\,J1228+1040, we detected photospheric Mg absorption, implying
ongoing accretion onto the white dwarf. As part of this study, we
obtained $I$-band spectra of WD\,1337+705, a well-known metal-polluted
DA white dwarf with a similar effective temperature as
SDSS\,J1043+0855 and SDSS\,J1228+1040, which did, however, not reveal
any Ca\,{\sc ii} emission. \citet{farihietal09-1} observed WD\,1337+705
with \textit{Spitzer} and did not detect any significant infrared
excess, and suggest that the detection of a dusty debris disk requires
an accretion rate of
$dM/dt\ge3\times10^8\mathrm{g\,s^{-1}}$. Similarly, the presence of
Ca\,{\sc ii} emission lines appears to correlate with the degree of metal
pollution. 

A third gaseous debris disk white dwarf was found in
DR\,6 \citep{gaensickeetal08-1}. SDSS\,J084539.17+225728.0 has a
helium-dominated (DB) atmosphere, which strongly limits the hydrogen content
of the circumstellar disk: because of the strong gravity of white
dwarfs, hydrogen will float on top of the atmosphere. The very low
hydrogen abundance in SDSS\,J0845+2257,
$\mathrm{H/He}\le3\times10^{-5}$, implies that the circumstellar
material is dramatically depleted in volatile elements with respect to
solar abundances, corroborating the ``rocky asteroid'' origin of the
debris disk. 

All three white dwarfs with gaseous debris disks have effective
temperatures in the range $\sim18000-22000$\,K, and masses which are
slightly above the mean mass for single white dwarfs, pointing towards
$\sim$A-type progenitor stars. 

\section{Outlook}
At the time of writing these proceedings, we pursue a number of
projects that will eventually lead to a better understanding of the
structure and origin of these gaseous debris disks. Analysing DR\,7,
we have identified two more systems, the DB white dwarf
SDSS\,J073842.56+183509.6 (independently found
by \citealt{dufouretal10-1} because of its very strong photospheric
metal lines~--~though these authors failed to note the presence of the
emission lines) and another DA white dwarf (Fig.\,1). The two new
additions extend the temperature where gaseous disks exist down to
$\sim13500$\,K. \textit{HST}/COS and UVES spectra of SDSS\,J1228+1040
and SDSS\,J0845+2257 have been obtained and will provide detailed
insight into the chemical abundances of the circumstellar
material. Spectroscopic monitoring of the Ca\,{\sc ii} emission lines
reveals changes in both the morphology of the profiles as well as of
the line fluxes, indicating that these debris disks evolve on time
scales of years~--~potentially suggesting that they are the hallmark
of recent disruption events that are settling down into a dust-only
configuration. Finally, a photoionisation model developed using
CLOUDY \citep{ferlandetal98-1} demonstrates that the observed line fluxes
can be understood by heating the top layers of the metal-rich gaseous
disks with ultraviolet photons from the white dwarf.

\begin{figure}
\centerline{\includegraphics[angle=270,width=0.98\textwidth]{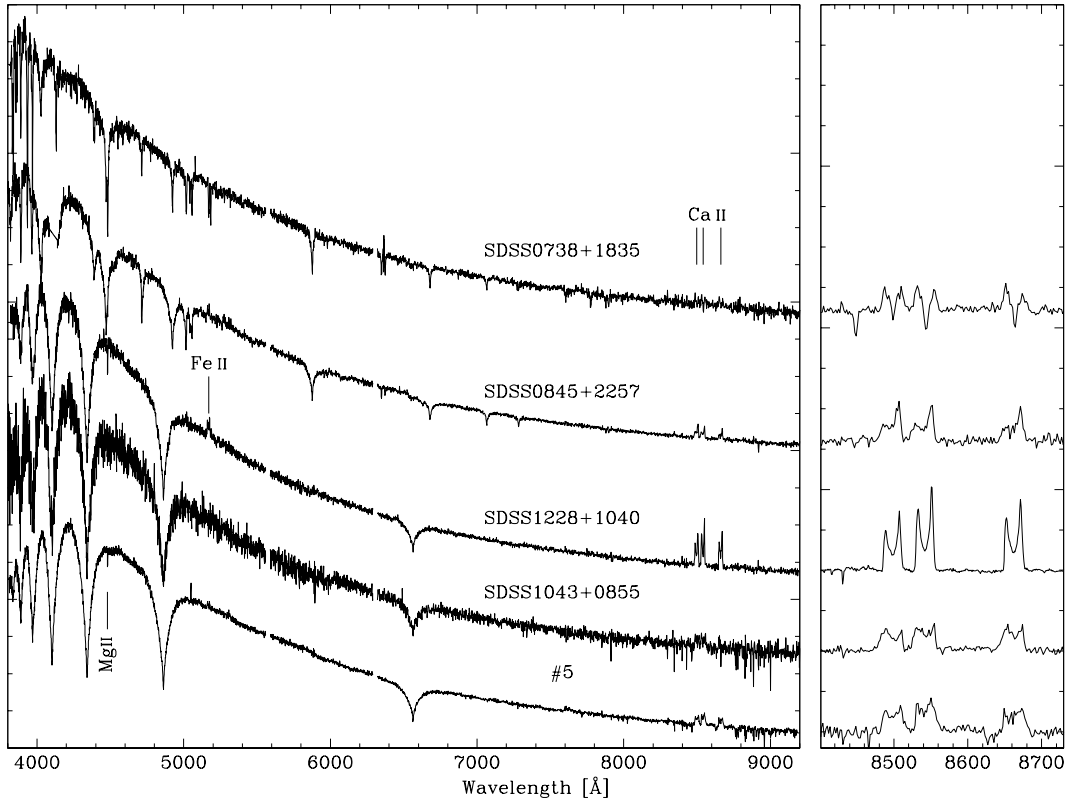}}
\caption{SDSS (left) and WHT (right) spectra of the five white dwarfs
with gaseous debris disks known so far
(from \citealt{gaensickeetal06-3, gaensickeetal07-1,
gaensickeetal08-1}, and G\"ansicke et al.\ in prep).}
\end{figure}

\begin{theacknowledgments}
 I would like to thank my collaborators on this topic for all their
 help and input over the past few years: Elm\'e Breedt, Carolyn
 Brinkworth, Jay Farihi, Jon Girven, Tim Kinnear, Detlev Koester, Tom
 Marsh, Stelios Pyrzas, Alberto Rebassa-Mansergas, John Southworth,
 and Claus Tappert.
\end{theacknowledgments}

\bibliographystyle{aipproc}   

\end{document}